\documentclass[aps,prb,secnumarabic, nobibnotes, twocolumn,superscriptaddress]{revtex4-1}

\usepackage{amsfonts}
\usepackage{mathrsfs}
\usepackage{amsmath}% needed for subequations
\usepackage{color}
\usepackage{natbib}
\usepackage{graphicx}
\usepackage{bm}% bold maths
\usepackage{amssymb}
\usepackage{xspace}
\usepackage{epstopdf}
\usepackage{dcolumn}% Align table columns on decimal point
\usepackage{longtable}
\usepackage{multirow}
\usepackage[colorlinks=true, letterpaper=true, pdfstartview=FitV, linkcolor=blue, citecolor=blue, urlcolor=blue]{hyperref}

\begin{document}

\title{Hourglass-like Nodal Net Semimetal in Ag$_2$BiO$_3$}
\author{Botao Fu}
\affiliation {Beijing Key Laboratory of Nanophotonics and Ultrafine Optoelectronic Systems, School of Physics, Beijing Institute of Technology, Beijing 100081 China}
\author{Dashuai Ma}
\affiliation {Beijing Key Laboratory of Nanophotonics and Ultrafine Optoelectronic Systems, School of Physics, Beijing Institute of Technology, Beijing 100081 China}
\author{Xiaotong Fan}
\affiliation {Beijing Key Laboratory of Nanophotonics and Ultrafine Optoelectronic Systems, School of Physics, Beijing Institute of Technology, Beijing 100081 China}
\author{Cheng-Cheng Liu}
\email{ccliu@bit.edu.cn}
\affiliation {Beijing Key Laboratory of Nanophotonics and Ultrafine Optoelectronic Systems, School of Physics, Beijing Institute of Technology, Beijing 100081 China}
\author{Yugui Yao}
\email{ygyao@bit.edu.cn }
\affiliation {Beijing Key Laboratory of Nanophotonics and Ultrafine Optoelectronic Systems, School of Physics, Beijing Institute of Technology, Beijing 100081 China}
%$k{\cdot}p$   % % ~\cite{zhu2015}
%%%%%%%%%%%%%%%%%%%%%%%%%%%%%%%%%%%%%%%% Nature 538, 75 (2016)
\begin{abstract}
Based on first-principles calculation and analysis of crystal symmetries, we propose a kind of hourglass-like nodal net (HNN) semimetal in centrosymmetric Ag$_2$BiO$_3$ that is constructed by two hourglass-like nodal chains (HNCs) at mutually orthogonal planes in the extended Brillouin zone (BZ) when the weak spin-orbit coupling (SOC) mainly from the 6s orbital of Bi atoms is ignored. The joint point in the nodal net structure is a special double Dirac point located at the BZ corner. Different from previous HNN [Bzdu{\v{s}}ek et al., Nature 538, 75 (2016)]
where the SOC and double groups nonsymmorphic symmetries are necessary, and also different from the accidental nodal net, this HNN structure is inevitably formed and guaranteed by spinless nonsymmorphic symmetries, and thus robust against any symmetry-remaining perturbations. The Fermi surface in Ag$_2$BiO$_3$ consisting of a torus-like electron pocket and a torus-like hole pocket may lead to unusual transport properties.
A simple four-band tight-binding model is built to reproduce the HNN structure.
For a semi-infinite Ag$_2$BiO$_3$, the ``drumhead'' like surface states with nearly flat dispersions are demonstrated on (001) and (100) surfaces, respectively.
If such weak SOC effect is taken into consideration, this HNN structure will be slightly broken, left a pair of hourglass-like Dirac points at the two-fold screw axis. This type of hourglass-like Dirac semimetal is symmetry-enforced and does not need band inversion anymore.
Our discovery provides a new platform to study novel topological semimetal states from nonsymmorphic symmetries.
\end{abstract}
\maketitle

\section{Introduction}
Beyond famous Dirac and Weyl semimetals\cite{armitage2017weyl,WengH2015rewtsm,HasanMZ2016top-rew,YanB2017rew,HirayamaMtop2018} which can be regarded as analogues of Dirac and Weyl fermions in particle physics,
a new kind of topological semimetal, nodal line semimetal (NLSM) ~\cite{burkov2011topological,Chour2017class,fang2015topological,fang2016topological,phillips2014tunable,yu2017topological,yang2017symmetry,takahashi2017spinless}, has recently been predicted and experimentally discovered~\cite{hu2016evidence,neupane2016observation,bian2016topological,feng2017experimental}.
For a NLSM, the lowest conduction and highest valence bands touch with each other at a series of continuous points, forming a closed loop in reciprocal space. The topological property of NLSM can be characterized by the accumulation of Berry phase~\cite{kim2015dirac} along a closed loop encircling the nodal line. According to bulk-boundary correspondence, a NLSM can exhibit a ``drumhead'' like surface state with nearly flat dispersion which may induce giant Friedel oscillations \cite{li2016dirac}, strong correlated effects and high temperature superconductivities~\cite{kopnin2011high}. With torus-like Fermi surface, a NLSM shows unusual transport properties such as multiple phase shifts in quantum oscillations\cite{yang2018quantum,rui2018topological}.

\begin{figure}
\includegraphics[width=3.5in]{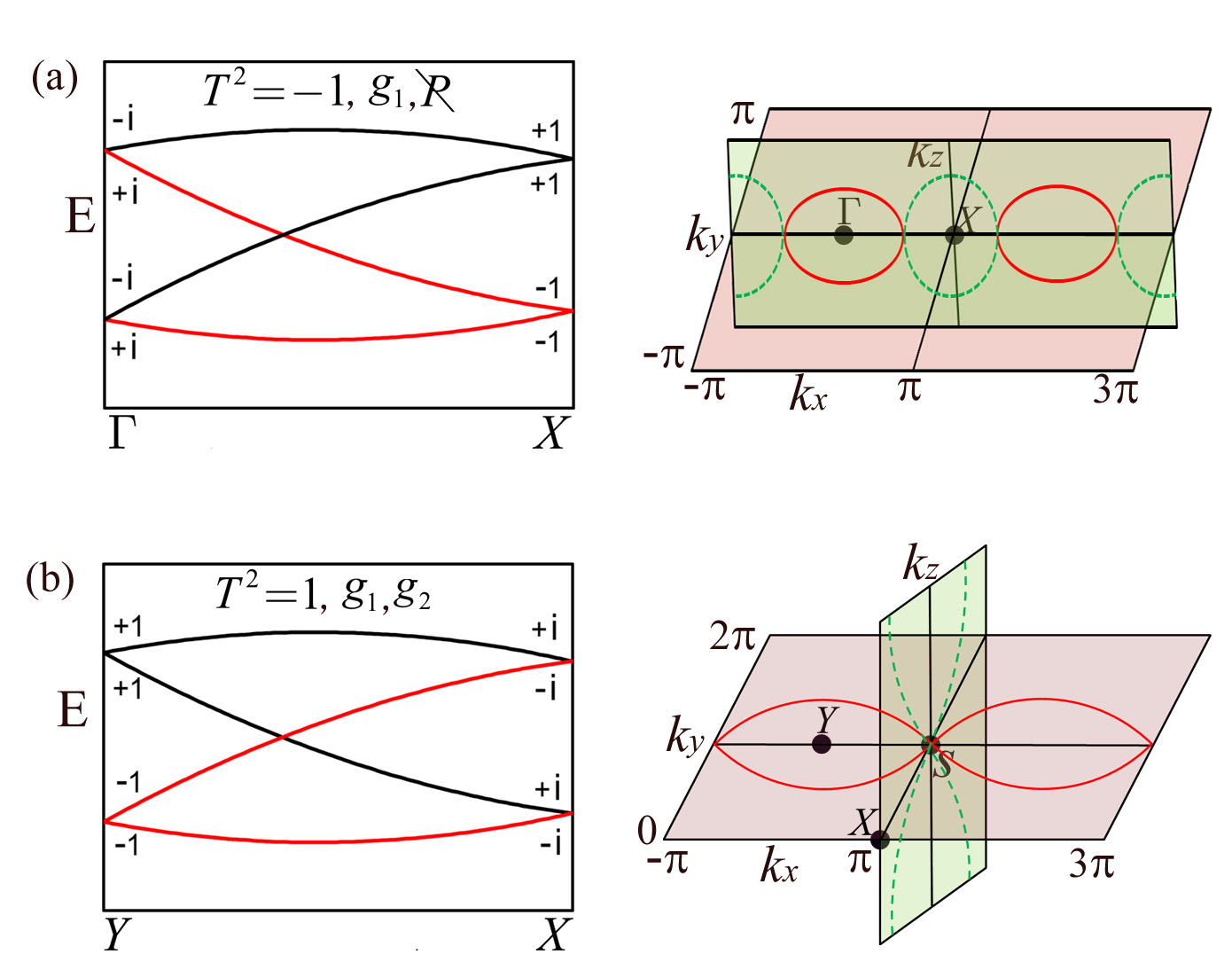}
\caption{(color online). (a) Left panel: the hourglass-like band structures protected by $g_1$=$\{{{M}_{z}}|\boldsymbol{a}/2\}$ for systems without $P$ symmetry including SOC effect. Right panel: the corresponding nodal rings at $k_x$$-$$k_y$ plane (red solid line) and $k_x$$-$$k_z$ plane (green dashed line) constructing an HNC in extended BZ.
(b) Left panel: hourglass-like band structures without SOC protected by $g_1$ and $g_2$ in the absence of SOC effect. Right panel: the corresponding HNCs at $k_x$$-$$k_y$ plane (red solid line) and $k_x$$-$$k_z$ (green dashed line) constructing an HNN structure in extended BZ.
 The eigenvalues of $\{{{M}_{z}}|\boldsymbol{a}/2\}$ are given. Notice that the HNC in (b) is formed by a large nodal line traversing the whole BZ while the HNC in (a) is formed by two tangent nodal rings alternately in orthogonal planes.}\label{schematic}
\end{figure}

%kim2015dirac,,,li2016dirac  robustness
According to the robustness, NLSMs can be divided into two categories: accidental nodal lines (ANLs) and symmetry-enforced nodal lines.
The ANLs are usually derived from band inversion mechanism and protected by the coexistence of time reversal and space inversion symmetries~\cite{yu2015topological,zhao2016topological,du2017cate,xu2017topological}
or mirror symmetry~\cite{yamakage2015line,xie2015new}.
This type of nodal lines can be annihilated without breaking the corresponding symmetry as long as the band inversion disappears.
On the other hand, a nonsymmorphic symmetry can protect the symmetry-enforced nodal line.
Suppose there is a slide mirror operation $g_1$=$\{{{M}_{z}}|\boldsymbol{a}/2\}$,
indicating a mirror reflection about $z$ direction accompanied by
a translation of half lattice vector $\boldsymbol{a}/2$.
On the mirror invariant plane (e.g. $k_z$=0), the momentum-dependent eigenvalues of $g_1$ are $\pm \lambda {{e}^{i{{k}_{x}/2}}}$, where ${\lambda}$=1 or $i$ for spinless or spinfull systems.
As shown in Fig. \ref{schematic} (a), for systems with time reversal symmetry (TRS) $T$ but without space inversion symmetry $P$,
all bands are generally splitting except at time reversal invariant momentums (TRIMs), where $T^2$=$-1$ ensures the Kramers degenerating.
Based on the evolution of this eigenvalues between two TRIMs (e.g. $\Gamma$ and $X$ points), two pairs of bands have to switch their partners and inevitably cross each other forming hourglass-like bandstructures~\cite{young2015dirac}.
Actually, along any loops connecting this two TRIMs in the plane, the band degenerate point always exists, which will form a closed nodal ring (red solid line) centered at $\Gamma$ point, as displayed in the right panel of Fig. \ref{schematic}(a).
Moreover, if the system has an additional glide operation on the perpendicular plane (e.g. $k{_x}-k{_z}$ plane), another nodal ring (green dashed line) will appear and two nodal rings touch each other to form an HNC structure~\cite{bzduvsek2016nodal,wang2017hourglassDNC}, as shown in the right panel of Fig. \ref{schematic}(a).
This symmetry-enforced nodal line and nodal chain were first put forward in Ref.~\cite{bzduvsek2016nodal}, and IrF$_4$ was proposed as a candidate material. However, this symmetry-enforced nodal lines in the material are subject to three constraints: remarkable SOC effect,  inversion symmetry broken, and the time-reversal symmetry preserved. The third constraint may be broken by the possible magnetic order in IrF$_4$ at low temperature. As a result, it is important and urgent to search for more and better symmetry-enforced nodal line semimetal candidates free of such constraints.

In this work, we find another path to realize symmetry-enforced nodal lines and net in weak SOC materials with spinless nonsymmorphic symmetries, which can be generalized to bosonic systems with similar symmetries.
As shown in the left panel of Fig. \ref{schematic}(b), in the absence of SOC, $T^2$=$1$ do not ensure Kramers degeneracy any more. However, the joint operation $\widetilde{T}$=${g}_{1}$$T$ satisfies $\widetilde{T}^{2}$=$-1$ at $k_x$=$\pi$, which can give equivalent Kramers degenerate at X point.
If we have another nonsymmorphic symmetry $g_2$ (e.g. a two-fold screw axis) that can ensure the double-degeneracy at $k_x$=0 (e.g. $Y$ point), then following the evolution of eigenvalues of $g_1$ from $Y$ to $X$, we deduce that a nodal line will appears in the $k_z$=0 plane.
In fact, as shown in the right panel of Fig. \ref{schematic}(b), rather than forming a nodal line, here the crossing point begins and ends at a four-fold degenerate point on the BZ corner (this will be discussed later), and finally forms an HNC structure (red solid line) in extended BZ.
If we further consider an extra perpendicular slide mirror plane, another HNC (green dashed line) emerges and two HNCs will link together forming a nodal net structure.
From this analysis, we propose that Ag$_2$BiO$_3$ with $Pnna$ space group can host an ideal HNN structure when the weak SOC effect is ignored.
The Fermi surface is made up of a torus-like electron pocket and a torus-like hole pocket which may induce novel transport properties.
After including SOC effect, the HNN is slightly gapped and a pair of hourglass-like Dirac points consequentially emerge, which are distinctive from previous Dirac semimetal protected by pure rotation symmetries~\cite{Yang2014Classification}.

%%%%%%%%%%%%%%%%%

\section{The calculation method and the Geometric structure of Ag$_2$BiO$_3$}

\begin{figure}
\includegraphics[width=3.5in]{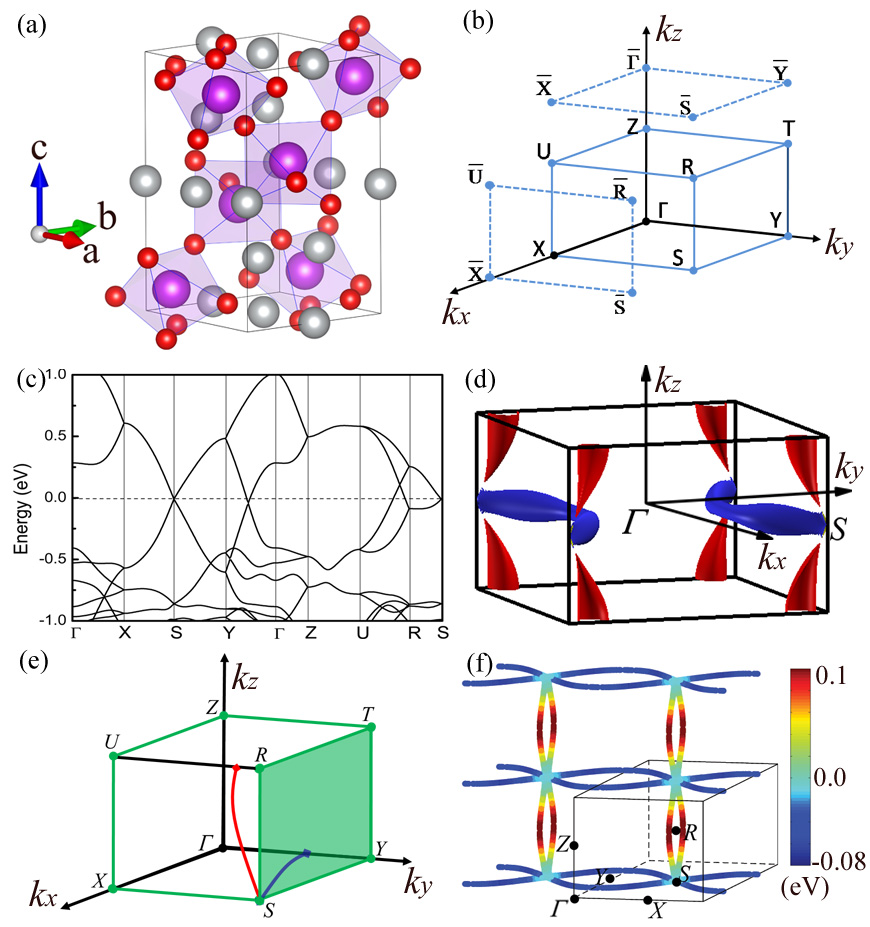}
\caption{(color online). (a) The unit cell of Ag$_2$BiO$_3$ in $Pnna$ phase. The violet, red and gray balls represent for bismuth, oxygen and silver atoms, respectively. (b) The bulk Brillouin zone and (001)/(100) surface Brillouin zone. (c) The band structure of Ag$_2$BiO$_3$ without SOC. (d) The torus-like Fermi surface with electron (red) and hole (blue) pockets. (e) Showing all the two-fold degenerate points (green color) and two nodal lines (red and blue color) in BZ. (f) Display of HNN structure in extended BZ. The color stands for the energy dispersion of HNN with respect to Fermi level.
}\label{dftband}
\end{figure}

Our first-principles calculations are performed by VASP (Vienna ab initio simulation package)~\cite{kresse1994ab,kresse1996efficient} within the generalized-gradient approximation (GGA)~\cite{blochl1994projector} of Perdew, Burke, and Ernzerhof (PBE)~\cite{perdew1996generalized}. The cutoff energy of 500 eV and the $k$-mesh of 15$\times$15$\times$15 are chosen to guarantee energy convergence. The crystal structure is relaxed to the ground states with force less than 0.01 eV/\AA\ on each atom. The surface states are obtained by using the method of maximally localized Wannier functions in Wannier90~\cite{marzari1997maximally,souza2001maximally} code and WannierTools code~\cite{wu2017wanniertools}.

The crystal structure of Ag$_2$BiO$_3$~\cite{deibele1999bismuth} used to be very controversial because of the discrepancy between theoretical predicted metallic behaviors and experimental observed insulating results. Subsequently, some careful analyses both in experiment and theoretical verified that Ag$_2$BiO$_3$ has three distinctive phases~\cite{oberndorfer2006charge,liu2007charge} with space groups of $Pnna$, $Pnn2$, and $Pn$, respectively. The $Pnna$ phase is a metal and the $Pn$ phase is a semiconductor, both of which are metastable and can transform into $Pnn2$ phase, which is a semiconducting ground state. In this work, we just discuss the $Pnna$ phase, and the formula Ag$_2$BiO$_3$ refers to the $Pnna$ phase unless specifically stated.

As shown in Fig. \ref{dftband}(a), the Ag$_2$BiO$_3$ crystallizes in a tetragonal crystal with centrosymmetric nonsymmorphic space group $Pnna$ (No. 52). There are four bismuth atoms (violet balls) in the unit cell that locate at the octahedral center of oxygen atoms (red balls) and eight sliver atoms (gray balls) fill the interstice sites. The optimized lattice constants are $|\boldsymbol{a}|$=6.13 {\AA}, $|\boldsymbol{b}|$=6.36 {\AA} and $|\boldsymbol{c}|$=9.83 {\AA}, which are consistent with experimental result~\cite{deibele1999bismuth}. The Brillouin zone and high symmetric points are displayed in Fig. \ref{dftband}(b).
%%%%%%%%%%%%%%%%%%%%%%%%%%%%%%%%%%%%%%%%%%%%%
\section{The electronic structure of Ag$_2$BiO$_3$}
The calculated electronic structure of Ag$_2$BiO$_3$ in the absence of SOC is shown in Fig. \ref{dftband}(c). We can see there are four bands entangling together near the Fermi level, which  are mainly derived from the 6$s$ orbital of bismuth atoms.
Besides, due to nonsymmorphic symmetries in $Pnna$ space group, both the conduction and valence bands possess two-fold degeneracy at high symmetric points (X, Y, R, S, etc.), lines (XS, SY, RS, etc.) and plane (RTYS) in the BZ, as shown in Fig. \ref{dftband}(c), (e). On one hand, along X-S-Y path, the lowest two-fold degenerate conduction and highest valence bands touch at S point at Fermi level and form a four-fold degenerate point.
In analogy with the concept of double Dirac point~\cite{wieder2016double,bouhon2017global} in the presence of SOC, we note this four-fold degenerate Dirac point in the absence of SOC as double Dirac point.
On the other hand, along Y$\Gamma$ and UR paths, two singlet bands cross each other, forming a single Dirac point located slightly below and above Fermi level, respectively. Remarkably, the band connection of four bands near Fermi level shows typically hourglass-like shape.

Moreover, through careful calculation, we discover that the trace of Dirac points actually forms a nodal line (NL) located in X$\Gamma$YS ($k_z$=0) and XURS ($k_x$=$\pi$) planes, respectively. Displayed in Fig. \ref{dftband}(e), the NL at XURS plane starts from S point, passes through UR and comes back to next S point in the extended BZ, which actually forms an HNC structure along SR direction as depicted in the right panel of Fig. \ref{schematic}(b).
On the other hand, the NL at X$\Gamma$YS plane also starts from S point, crosses ${\Gamma}$Y and forms another HNC structure along SY direction.
Those two HNCs joint together at the double Dirac point S, and thus forms an unique HNN structure as plotted in Fig. \ref{dftband}(f).
It is significant to find that the HNCs constructing the HNN structure have different energies. One HNC on $k_z$=0 plane has almost negative energy while the other on $k_x$=$\pi$ plane has nearly positive energy with respect to the Fermi level.
By plotting the Fermi surface in Fig. \ref{dftband}(d), we find a torus-like electron pocket (blue color) originated from the HNC at $k_z$=0 plane and a hole pocket (red color) originated from the HNC at $k_x$=$\pi$ plane, respectively. From this perspective, two HNCs constructing HNN are separated both in momentum space and in energy space except around S point.
So far as we know, the HNN structure proposed in this paper is different from all known nodal line, nodal chain and nodal net\cite{sun2017dirac,Yu2017From,feng2017topological} systems, we hope this unique HNN structure would result in novel properties in future transport experiments.

\section{Symmetry analysis of Ag$_2$BiO$_3$}

As mentioned before, the emergence of HNN semimetal in Ag$_2$BiO$_3$ without SOC results from specific nonsymmorphic symmetries. Based on symmetries analysis, we are going to prove the existence of HNN structure in Ag$_2$BiO$_3$  with the space group of $Pnna$.
The $Pnna$ space group totally includes eight symmetry operations: $I$, $P$, $\{{{C}_{2x}}|\boldsymbol{b}/2+\boldsymbol{c}/2\}$, $\{{{C}_{2z}}|\boldsymbol{a}/2\}$,$\{{{C}_{2y}}|\boldsymbol{a}/2+\boldsymbol{b}/2+\boldsymbol{c}/2\}$, $\{{{M}_{x}}|\boldsymbol{b}/2+\boldsymbol{c}/2\}$, $\{{{M}_{y}}|\boldsymbol{a}/2+\boldsymbol{b}/2+\boldsymbol{c}/2\}$, $\{{{M}_{z}}|\boldsymbol{a}/2\}$, where the first is the identity operation, the second is the inversion operation, the third and the fourth are two-fold rotations along $x$ and $z$ axes, respectively. The last four are effective nonsymmorphic operations.

Firstly, to prove the double degeneracy of the band structure at BZ boundaries, we consider a nonsymmorphic operation ${{g}_{1}}$=$\{{{M}_{z}}|\boldsymbol{a}/2\}$ and TRS $T$, which jointly act on lattice momentum as follows:
\begin{align}
  & {{g}_{1}}:\text{ }({{k}_{x}},{{k}_{y}},{{k}_{z}})\text{ }\to \text{ }({{k}_{x}},{{k}_{y}},-{{k}_{z}}), \\
  & T:\text{                 }({{k}_{x}},{{k}_{y}},-{{k}_{z}})\text{ }\to \text{ }(-{{k}_{x}},-{{k}_{y}},{{k}_{z}}).
\end{align}
Defining a joint operation $\widetilde{T}$=${g}_{1}$$T$, we find it commutes with
the Hamiltonian $H(\boldsymbol{k})$ when $\boldsymbol{k}$=(0/$\pi$, 0/$\pi$, $k_z$).
On the other band, with the conditions of ${{T}^{2}}$$\equiv$1 and $g_{1}^{2}$=${{e}^{i{{k}_{x}}}}$, we can get ${{{\widetilde{T}}}^{2}}$=$-1$ when ${{k}_{x}}$=$\pi$.
As a result, the Kramers-like band degeneracy~\cite{young2015dirac} can be obtained along XU ($\pi$, 0, $k_z$) and RS ($\pi$, $\pi$, $k_z$) paths. Similarly, the operation $g_2$=$\{{{M}_{y}}|\boldsymbol{a}/2+\boldsymbol{b}/2+\boldsymbol{c}/2\}$ guarantees band degeneracy along ZT and XS paths, and the operation $g_3$=$\{{{M}_{x}}|\boldsymbol{b}/2+\boldsymbol{c}/2\}$ enforces band degeneracy along ZU and YS paths. Besides, the two-fold screw axis  $g_4$=$\{{{C}_{2y}}|\boldsymbol{a}/2+\boldsymbol{b}/2+\boldsymbol{c}/2\}$ enforces band degeneracy in the whole RTYS (${{k}_{y}}$=$\pi$) plane. Those degenerating points in the BZ are clearly demonstrated in Fig. \ref{dftband}(e).

Secondly, to prove the four-fold degeneracy at S ($\pi$, $\pi$, 0) point,
we chose three symmetry operations: $P$, $g_2$ and $g_5$=$\{{{C}_{2z}}|\boldsymbol{a}/2\}$
that all commute with $H(\boldsymbol{k})$ at S point. It is easy to obtain the relations: ${P}^{2}$=1, $g_{2}^{2}$=$-1$ and $g_{5}^{2}$=1.
In analogy with above analysis, combining $g_{2}^{2}$=$-1$ with $T$, we can acquire two Kramers-like degenerate states,
such as $|\phi_{+i}^{E}{\rangle}$ and $T|\phi _{+i}^{E}{\rangle}$ with eigenvalues $+i$ and $-i$ of $g_2$, respectively. Utilizing the commutation relation $[P,{{g}_{2}}]$=0, we can additionally label those states by the parity $\lambda$ as $|\phi_{\lambda ,+i}^{E}{\rangle}$, $T|\phi_{\lambda ,+i}^{E}{\rangle}$. Since $T$ won't change parity, those two states must have same parity.
On the other band, using the anti-commutation relation $\{P,{{g}_{5}}\}$=0, we can deduce that
$|\phi _{\lambda ,+i}^{E}{\rangle}$, ($T|\phi _{\lambda ,+i}^{E}{\rangle}$)
must have its degenerate partner ${g}_{5}|\phi _{\lambda ,+i}^{E}{\rangle}$, (${{g}_{5}}T|\phi _{\lambda ,+i}^{E}{\rangle}$) with opposite parity. Therefore, we obtain the four-fold degeneracy of Bloch states at S point.

\begin{figure}
\includegraphics[width=3.5in]{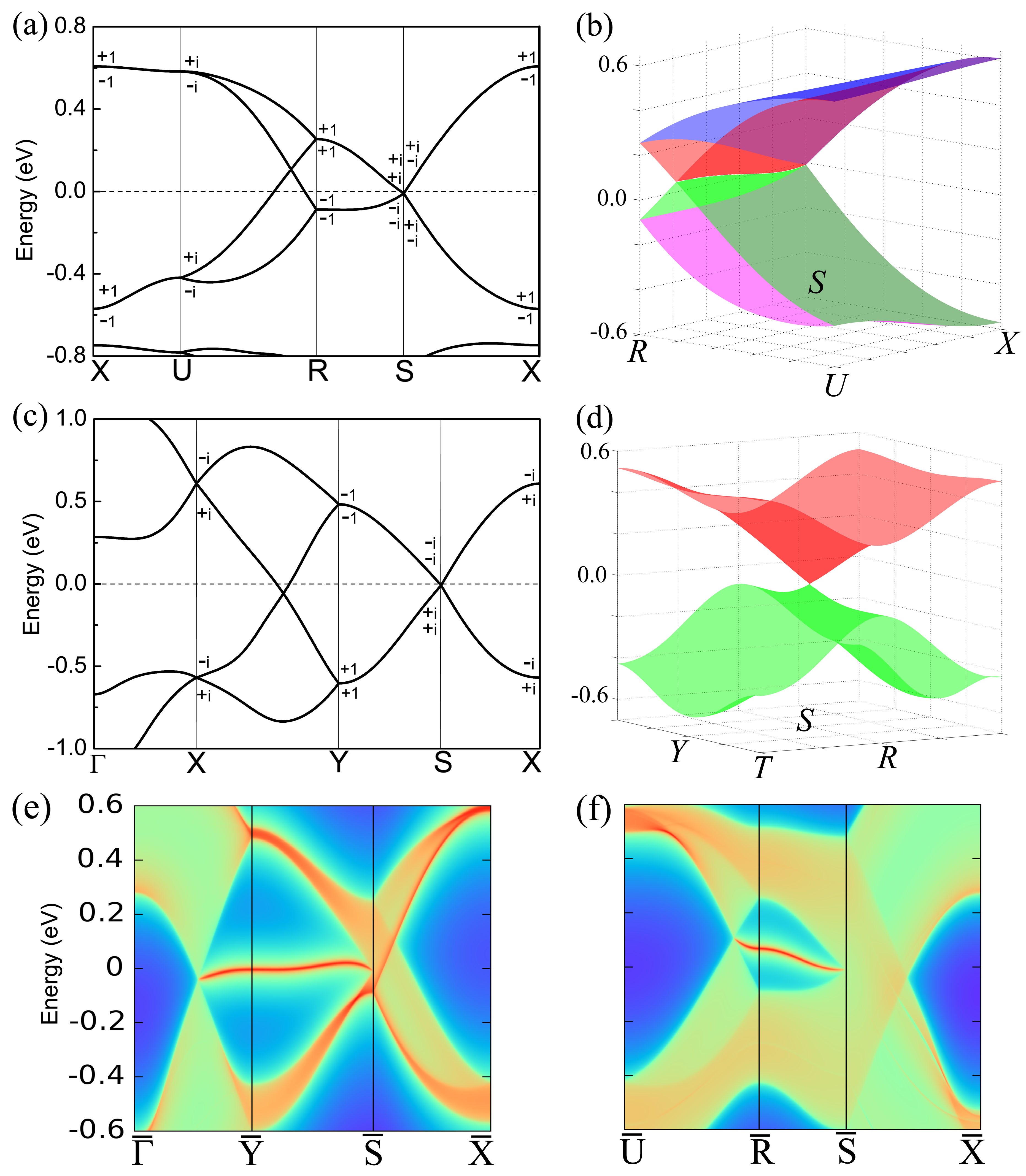}
\caption{(color online). (a) The band structure at XURS plane. The eigenvalues of $\{{{M}_{x}}|\boldsymbol{b}/2+\boldsymbol{c}/2\}$ are shown. (b) 3D view of energy spectrum at XURS plane. (c) The band structure at X$\Gamma$YS plane. The eigenvalues of $\{{{M}_{z}}|\boldsymbol{a}/2\}$ are shown. (d) 3D view of double Dirac cone around S point at RTYS plane. The bands are doubly degenerate. (e), (f) Calculated surface states of Ag$_2$BiO$_3$ on (001) and (100) surfaces, respectively.
}\label{symmetry}
\end{figure}

Keeping those in mind, we now discuss the band connections and corresponding band crossings at each mirror invariant planes.
For $g_3$=$\{{{M}_{x}}|\boldsymbol{b}/2+\boldsymbol{c}/2\}$ operation, the invariant plane is URSX ($k_x$=$\pi$), we can use its eigenvalues ${{g}_{\pm }}({{k}_{y},{k}_{z}})$=$\pm {{e}^{i{{k}_{y}/2+i{k}_{z}/2}}}$ to identify each bands.
Specifically, at U ($\pi$, 0, $\pi$) and S ($\pi$, $\pi$, 0), a pair of degenerate bands have opposite eigenvalues ($\pm i$), which are interchanged by $T$. While at X ($\pi$, 0, 0) and R ($\pi$, $\pi$, $\pi$) points, the eigenvalues can take $+1$ or $-1$. Because the bands are doubly degenerate along UX path, the eigenvalues for a pair of degenerate bands will evolve from $\pm i$ at U point to $\pm 1$ at X point.
Therefore, the quadruple degenerate states at S point with eigenvalues ($+i,-i,+i,-i$) will split into two pairs of double-degenerate states along SX path with eigenvalues ($+i,-i$) and ($+i,-i$), respectively, as shown in Fig. \ref{symmetry}(a).
On the other hand, since the degeneracy along SR path is protected by $g_4T$ that commutes with $g_3$ at R point, we can easily derive that the double-degenerate bands at R point have the same eigenvalues ($+1, +1$) or ($-1, -1$).
Hence, the quadruple degenerate states at S point with eigenvalues ($+i,-i,+i,-i$) will separate into two pairs of double-degenerate states along SR path with eigenvalues ($+i,+i$) and ($-i,-i$), respectively, as shown in Fig. \ref{symmetry}(a).
As a consequence, along UR path, two pairs of bands have to switch partners and cross each other, forming a typical hourglass-like structure. Along arbitrary loops connecting U and R points in this plane, the crossing point is always existing and its trace naturally forms one quarter of nodal line in the plane as shown in Fig. \ref{symmetry}(b).
A significant difference between this type of nodal line (Fig. \ref{schematic}(b)) and other nodal lines (Fig. \ref{schematic}(a)) is that there exist a double Dirac point S. The nodal line starts from S point, crosses UR line and comes back to another S point, which forms a nodal chain in extended BZ. It's worth noting that the double Dirac point plays a crucial role in forming the nodal chain structure.

Similarly, for $g_1$=$\{{{M}_{z}}|\boldsymbol{a}/2\}$ operation at its invariant plane X$\Gamma$YS ($k_z$=0), using the commutation relation of $g_4$ and $g_1$,
we obtain that two degenerate bands possess the same eigenvalues at Y point and opposite eigenvalues at X point as shown in Fig. \ref{symmetry}(c). Thus, an HNC is also expected in this plane. Finally, the HNCs in two mutually perpendicular planes link together at S point forming the HNN structure as demonstrated in Fig. \ref{dftband}(f).
In addition, at SRTY ($k_y$=$\pi$) plane, both conduction bands and valence bands are
double-degenerate due to $g_4$=$\{{{C}_{2y}}|\boldsymbol{a}/2+\boldsymbol{b}/2+\boldsymbol{c}/2\}$ operation.
They touch each other at S point, forming a double Dirac cone with linearly dispersion as shown in Fig. \ref{symmetry}(d).

\section{The surface states of Ag$_2$BiO$_3$}
Based on the maximally-localised Wannier functions (MLWF) methods~\cite{souza2001maximally,wu2017wanniertools},  we calculate the surface energy spectrum on different surfaces as shown in Fig. \ref{symmetry}(e) and (f). For (001) surface, the HNC on $k_z$=0 plane projects into this surface.  As in Fig. \ref{symmetry}(e), we can see inside the projected region there are ``drumhead'' like surface states with nearly flat dispersion while there are no surface states outside the projected region, which indicates the topological property of HNC in the bulk state. The other HNC at $k_x$=$\pi$ plane projects into $\bar{\text{S}}$$\bar{\text{X}}$ line and gives only bulk states. For (100) surface, the HNC at $k_x$=$\pi$ plane will project into this surface, and similar flat surface state emerges as shown in Fig. \ref{symmetry}(f). The HNC at $k_z$=0 projects into $\bar{\text{S}}$$\bar{\text{X}}$ line and only gives bulk states. The region of projected HNC is rather sizable in momentum space and the surface state is close to Fermi level, which facilitates the observations in experiment.

\section{tight-binding model of Ag$_2$BiO$_3$}

\begin{figure}
\includegraphics[width=3.5in]{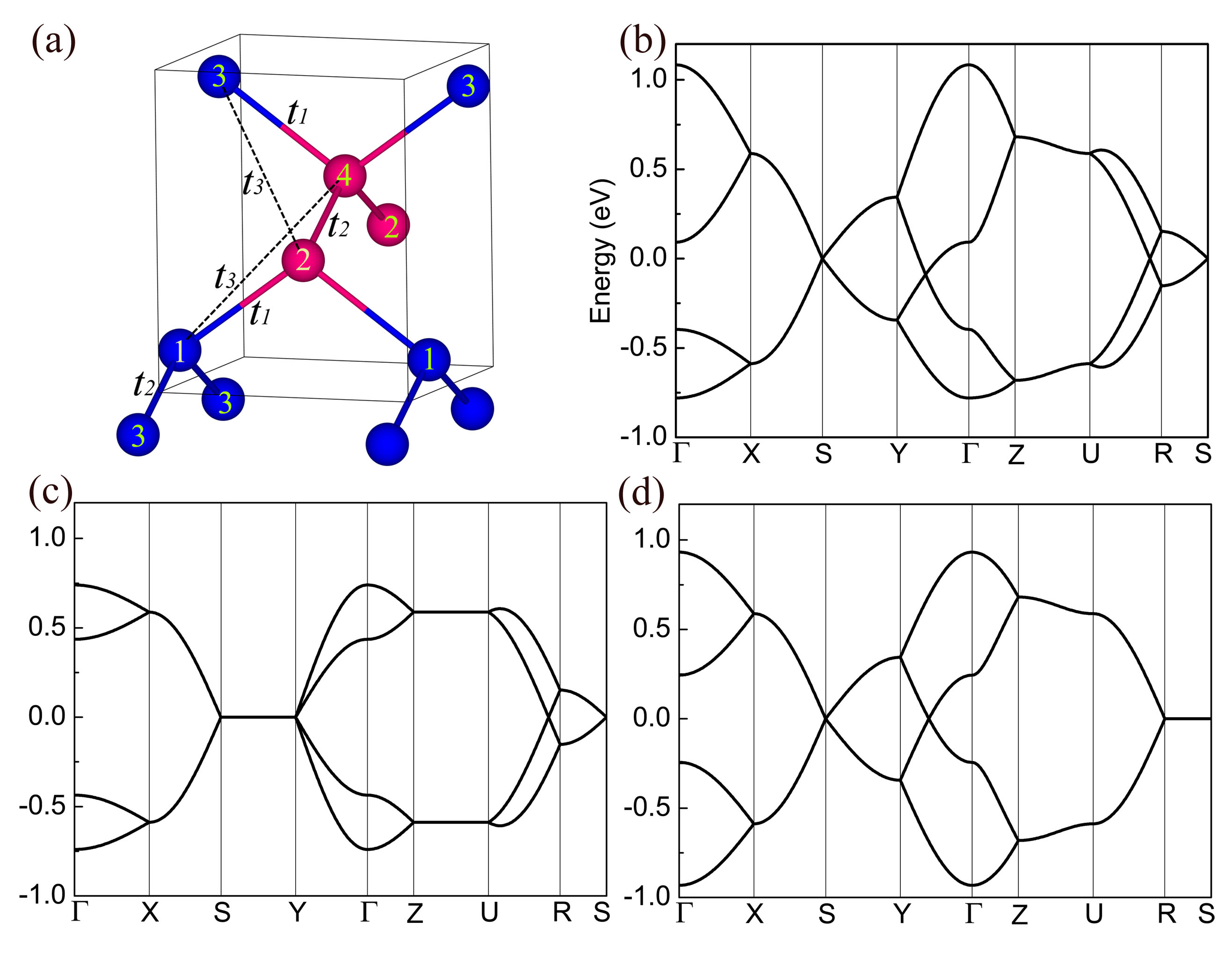}
\caption{(color online). (a) The lattice with four bismuth atoms for TB model. $t_1$ is the NN hopping between site 1 and 2, 3 and 4.
$t_2$ is the NN hopping between site 2 and 4, 1 and 3. $t_3$ is the NNN hopping between 1 and 4, 2 and 3. (b) TB band structure ($t_1$=-0.294 eV, $t_2$=-0.172 eV, $t_3$=0.038 eV, ${{\varepsilon }_{i}}$=0.0 eV). (c) TB band structure with the same parameters as (b) except $t_2$=0.0 eV. (d) TB band structure with the same parameters as (b) except $t_3$=0.0 eV.
}\label{tbband}
\end{figure}

Considering that the electronic states near the Fermi level mainly originate from $6s$ orbital of four bismuth atoms, we can use a four-band ting-binding (TB) model to describe the band structure of Ag$_2$BiO$_3$. The base set is chosen as $\{{{\phi }_{i,s}}\}$, where $i$=1, 2, 3, 4 stands for different sites as shown in Fig. \ref{tbband} (a).
The TB model is written as
\begin{align}
&H=\sum\limits_{i}{{{\varepsilon }_{i}}c_{i}^{\dagger }{{c}_{i}}}+\sum\limits_{i\ne j}{{{t}_{ij}}c_{i}^{\dagger }{{c}_{j}}},
\end{align}
where $c_{i}^{\dagger }({{c}_{j}})$ is the creation (annihilation) operator of electrons at site $i$ ($j$), ${{t}_{ij}}$ is the hopping parameter between the $i$th and $j$th atoms, ${{\varepsilon }_{i}}$ is the on-site energy of $i$th site.

As shown in Fig. \ref{tbband}(a), we consider two nearest neighbor (NN) hoppings ($t_1$, $t_2$) and one next nearest neighbor (NNN) hopping ($t_3$). By fitting those hopping parameters, we find the band structure from TB model matches well with the DFT result. Especially, it can well reproduce the HNN structure in Fig. \ref{tbband}(b).
In addition, we reveal that the NN hopping $t_2$ is responsible for the forming of HNC on X$\Gamma$YS plane.
In Fig. \ref{tbband}(c), when $t_2$ gradually decreases to zero, the HNC on $\Gamma$YSX plane will be compressed and finally transform into a four-fold degenerate nodal line along SY path, while the HNC on URXS plane is maintained.
The HNC on URXS plane is dominated by NNN hopping, it will transform into a four-fold degenerate nodal line along RS path when $t{_3}$=$0$, as shown in Fig. \ref{tbband}(d). This transformation from HNC to nodal line is accompanied with the change of symmetry.

\section{SOC effect on the HNN}
In the presence of SOC, bands are conventionally Kramers degenerate due to the coexistence of time reversal and space inversion symmetries.
A pair of Kramers degenerating states at each $k$ point can generally be written as $|\phi {\rangle}$ and $PT|\phi {\rangle}$.
To determine whether the HNN structure can still exist, we need to recalculate the eigenvalues of corresponding symmetries for all bands under SOC condition.

\begin{figure}
\includegraphics[width=3.4in]{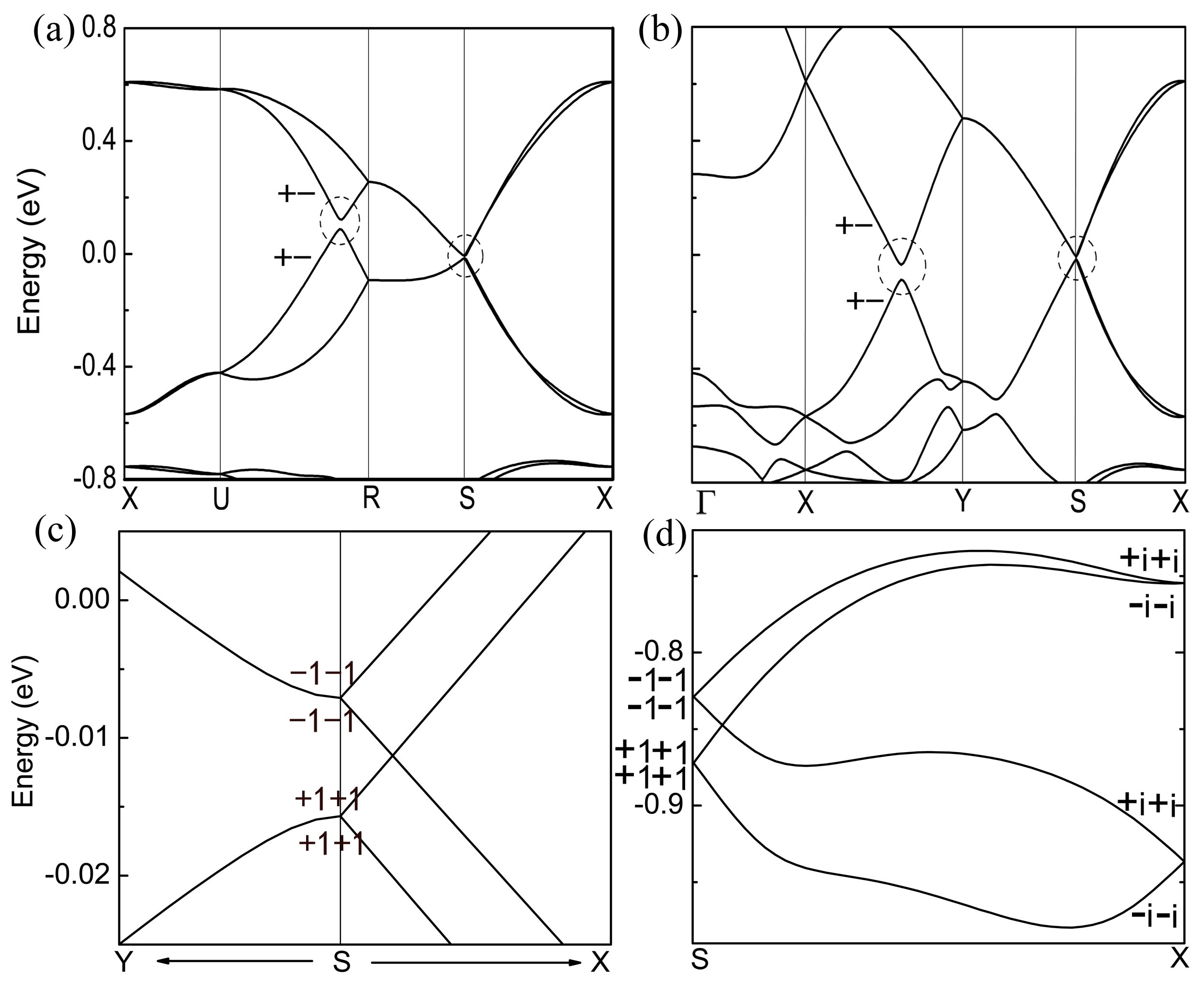}
\caption{(color online). (a), (b) Band structures from DFT including SOC. There are gaps inside the dashed circles. (c) The enlarged band structure around S point near Fermi level. (d) The hourglass-like band structure along SX below Fermi level. The eigenvalues of $\{{{C}_{2y}}|a/2+b/2+c/2\}$ are given.
}\label{socband}
\end{figure}

At $k_x$=$\pi$ plane, we can use the eigenvalues ${{g}_{\pm }}$=$\pm i{{e}^{i({{k}_{y}}/2+{{k}_{z}}/2)}}$ of $g_3$=$\{{{M}_{x}}|\boldsymbol{b}/2+\boldsymbol{c}/2\}$
to identify each bands.
Suppose a state $|\phi {\rangle}$ has eigenvalue ${g}_{+}$, which satisfies
${{g}_{3}}|{{\phi }}{\rangle}=+i{{e}^{i({{k}_{y}}/2+{{k}_{z}}/2)}}|{{\phi }}{\rangle}$.
Using the commutation relation ${{g}_{3}}P={{e}^{i{{k}_{y}}+i{{k}_{z}}}}P{{g}_{3}}$,
we find the Kramers partner $PT|\phi {\rangle}$ satisfies
${{g}_{3}}PT|\phi {\rangle}={{e}^{i{{k}_{y}}+i{{k}_{z}}}}PT{{g}_{3}}|\phi {\rangle}=-i{{e}^{i({{k}_{y}}/2+{{k}_{z}}/2)}}PT|\phi {\rangle}$,
which indicates $PT|\phi {\rangle}$ has eigenvalue ${{g}_{-}}$. Therefore, we obtain that a pair of Kramers degenerate bands must have opposite eigenvalues of $g_3$ as shown in Fig. \ref{socband}(a).
Consequently, if two pairs of bands touch each other, $g_3$ cannot protect a four-fold degeneracy and gaps are generally opened between bands with the same eigenvalues.
Through similar symmetry analysis, we find $g_1$=$\{{{M}_{z}}|\boldsymbol{a}/2\}$ cannot protect a four-fold degeneracy at $k_z$=0 plane. Thus, we conclude that the HNCs (without SOC) on two planes will be gapped out by the SOC effect.
As shown in Fig. \ref{socband}(a) and (b), because of the relatively weak SOC effect of $s$ orbitals, the band gaps induced by SOC are about 51 meV along UR, 33 meV along XY and 8 meV at S point, respectively.
%Therefore, the HNN structure is in fact broken by SOC, even though the gaps are relatively small.

On the other hand, along SX path, we discover a pair of conduction bands switch one partner with a pair of valence bands and inevitably forms an
hourglass-like Dirac semimetal (HDSM)~\cite{wieder2016spin} structure as shown in Fig. \ref{socband}(c). This process happens for every group of four Kramers pairs in all energy range displayed in Fig. \ref{socband}(d).
To reveal the formation of HDSM, we focus on ${{g}_{4}}$=$\{{{C}_{2y}}|a/2,b/2,c/2\}$ operation acting on Bloch states at SX path.  We can use its eigenvalues ${{g}_{\pm }}$=$\pm i{{e}^{i{{k}_{y}}/2}}$ to label all bands.
Utilizing the condition ${{g}_{4}}P=-{{e}^{i{{k}_{y}}}}P{{g}_{4}}$,
we can verify that two Kramers degenerate bands must have the same eigenvalues of $g_4$ along SX.
Therefore, in Fig. \ref{socband}(d), a pair of Kramers bands with eigenvalues ($+i,+i$) at X point
must be degenerate with a pair of bands with eigenvalues ($-i,-i$) since $T$ can interchange $i$ and $-i$ states.
Then, for the Bloch states at S point, a four-fold degeneracy can be derived by the anti-commutation relation $\{g_{1},P\}$=0, analogously to the proof of the four-fold degeneracy at S point in the above spinless case.
Moreover, because of that $g_4$ commutes with both $g_1$ and $P$ at S point, we can conclude that a four-fold degenerate Bloch states must have same eigenvalues of $g_4$, as shown in Fig. \ref{socband}(d).
Finally, along SX path, four pairs of Kramers degenerate bands have to switch partners and inevitably cross each other,
forming an hourglass-like dispersion relation.
There is a symmetry-enforced Dirac point that cannot be eliminated as long as the symmetry is remained, which is distinctive from the Dirac semimetals protected by pure rotation symmetries that only appearing during topological phase transitions~\cite{Murakami2007Phase}. If the electron filling number is $8n+4$~\cite{takahashi2017spinless}, the Dirac point of HDSM can perfectly cross the Fermi level in principle.
%Unlike the hourglass fermion in surface states~\cite{Wang2016Hourglass}, the hourglass Dirac fermion here are in the 3D bulk states and composed of four double-degenerated bands, which may provide new properties to be revealed in the future.

\section{Conclusions and discussions}

In summary, we first report an HNN structure in Ag$_2$BiO$_3$ with centrosymmetric nonsymmorphic space group in the absence of SOC.
The HNN structure is constructed by two HNCs, which are symmetry-enforced and cannot be removed without breaking symmetries.
From this perspective, this kind of nodal net is different from the accidental nodal net in  AlB${_2}$-type TiB$_2$\cite{feng2017topological} which is protected by spatial-inversion, mirror symmetry and band inversion, and can be annihilated by just lifting band positions.
The special torus-like Fermi surface at both electron and  hole sides are specifically shown.  A four-band tight-bind model is developed to describe the HNN structure and corresponding ``drumhead'' like surface state are demonstrated on (100) and (001) surfaces, respectively. In addition, when including SOC effect, the HNN are slightly gapped and a pair of Dirac points with hourglass-like dispersions inevitably emerge on a two-fold screw axis.

It is known that an electron can pick up a nontrivial $\pi$ Berry phase around a loop that interlocks with
the nodal line which may be probed by the Shubnikov-de Haas (SdH) quantum oscillations in experiment\cite{mikitik1999Man}.
It has been reveal\cite{Li2018Rules} that the total phase shift for each frequency component of the quantum oscillation depends on the extreme cross sections of Fermi surface, the direction of magnetic field and the sign of charge carrier. In this respect,  it is worth looking forward to observing the abundant phase shift patterns in Ag$_2$BiO$_3$ in quantum oscillation experiments, due to the unique Fermi surface made of a torus-like electron pocket and a torus-like hole pocket. Besides, we suggest this HNN structure can exist in weak SOC compounds or bosonic systems with similar nonsymmorphic symmetries.
%Besides, our symmetry analysis for formation of the HNN structure is effective for other spinless systems and can be generalized to bosonic systems.

\begin{acknowledgments}
This work was supported by the National Key R{\&}D Program of China (No. 2016YFA0300600), the National Natural Science Foundation of China (Nos. 11774028, 11734003, 11574029, 11404022),  the MOST Project of China (No. 2014CB920903), and Basic Research Funds of Beijing Institute of Technology (Grant No. 2017CX01018).

\end{acknowledgments}

\bibliography{Refer}

\end{document}